\documentclass[runningheads]{llncs}
\usepackage{graphicx}
\usepackage{listings}
\usepackage{subfig}
\usepackage{amsmath}
\usepackage{wrapfig}
\usepackage{bm}
\usepackage{marvosym}

\usepackage{amsfonts}
\usepackage{stmaryrd}
\usepackage{amssymb}

\usepackage{pifont}

\usepackage{mathtools}
\usepackage{mathpartir}
\usepackage{tabularx}
\usepackage[all,cmtip]{xy}
\usepackage{diagbox}
\newcolumntype{C}{>{\centering\arraybackslash}X}
\usepackage{hyperref}

\newcommand{\C}[1]{\texttt{#1}}
\newcommand{\F}[1]{\mathsf{#1}}

\newcommand{\event}{\mathsf{E}}
\newcommand{\invo}{\mathsf{I}}
\newcommand{\val}{\mathsf{V}}
\newcommand{\sess}{\mathsf{S}}

\newcommand{\rulelabel}[1]{\textrm{\sc {#1}}}

\newcommand{\ALT}{~\mid~}
\newcommand{\RULE}[2]{\frac{\begin{array}{c}#1\end{array}}
                           {\begin{array}{c}#2\end{array}}}

\begin{document}
\title{Semantics, Specification, and Bounded Verification of Concurrent Libraries in Replicated Systems}
\titlerunning{Bounded Verification of Concurrent Libraries in Replicated Systems}
%
\author{Kartik Nagar\inst{1} \and Prasita Mukherjee\inst{2} \and Suresh Jagannathan\inst{3}}
\institute{IIT Madras, India. \email{nagark@cse.iitm.ac.in} \and Purdue University, USA. \email{mukher39@purdue.edu} \and Purdue University, USA. \email{suresh@cs.purdue.edu}}
%

%
\maketitle              
\begin{abstract}

  Geo-replicated systems provide a number of desirable properties such
  as globally low latency, high availability, scalability, and
  built-in fault tolerance.  Unfortunately, programming correct
  applications on top of such systems has proven to be very
  challenging, in large part because of the weak consistency
  guarantees they offer.  While a large number of consistency policies
  have been proposed in recent years to aid programmers in developing
  applications suitable for these environments, profitably balancing
  correctness and efficiency by identifying the weakest policy under
  which an application can run correctly remains a highly non-trivial
  endeavor.\\[-2mm]

  These complexities are exacerbated when we try to adapt existing
  highly-performant concurrent libraries developed for shared-memory
  environments to this setting.  The use of these libraries, developed
  with performance and scalability in mind, is highly desirable.  But,
  identifying a suitable notion of correctness to check their validity
  under a weakly consistent execution model has not been well-studied,
  in large part because it is problematic to na\"ively transplant
  criteria such as linearizability that has a useful interpretation in
  a shared-memory context to a distributed one where the cost of
  imposing a (logical) global ordering on all actions is prohibitive.\\[-2mm]

  In this paper, we tackle these issues by proposing appropriate
  semantics and specifications for highly-concurrent libraries in a
  weakly-consistent, replicated setting.  We use these specifications
  to develop a static analysis framework that can automatically
  detect correctness violations of library implementations
  parameterized with respect to the different consistency policies
  provided by the underlying system.  We use our framework to analyze
  the behavior of a number of highly non-trivial library
  implementations, including stacks, queues, and exchangers.  Our
  results provide the first demonstration that automated correctness
  checking of concurrent libraries in a weakly geo-replicated setting
  is both feasible and practical.

\end{abstract}

\section{Introduction}

Geo-replicated systems maintain multiple copies of data at different
locations and provide a number of attractive properties such as
globally uniform low access-latency, always-on availability, fault
tolerance, and improved scalability.  Applications with a
geo-distributed user base need to necessarily run on top of replicated
systems to ensure fast and always-available service. On the other
hand, due to concurrent updates at different replicas and the
possibility of arbitrary re-ordering of updates by the underlying
network, replicated systems typically guarantee a very weak form of
consistency called \emph{eventual consistency}~\cite{BA13}, that only
requires replicas which have received the same set of updates to
exhibit the same state.  Because this guarantee is often too weak to
satisfy an application's correctness requirements, a number of
(stronger) consistency policies have emerged in recent years; these
policies offer session\cite{TDP94}, causality\cite{LL11} or
transactional \cite{CBG15} guarantees, and constrain system behavior
by imposing additional synchronization on actions. Nonetheless,
writing correct applications in this environment using these policies
remains a challenging problem.

Having a library of performant \emph{and correct} data structure
implementations developed with replication and geo-distribution in
mind can significantly alleviate the problem of writing correct
applications, as demonstrated by the availability of highly popular
concurrent library implementations developed for shared-memory
systems~\cite{PGB+05,HN08}.  CRDTs~\cite{SH11a} (Conflict-Free
Replicated Data Types) offer an analog of such implementations for
geo-replicated environments.  However, using CRDTS to build useful
data structure libraries is challenging because the strong
requirements imposed by CRDTs (namely that all operations commute with
each other) appears satisfiable only for simple objects such as sets,
lists, or maps.  Important data structures such as stacks, queues, or
exchangers that serve as building blocks for many concurrent and
distributed algorithms have eluded implementations using CRDTs.  Even
when a data structure can be expressed in this way, reasoning about
its correctness is typically given in terms of non-standard criteria
such as replicated data type specifications\cite{BU14},
convergence\cite{NJ19} or replication-aware
linearizability\cite{WEMP19}, concepts that are likely to be difficult
for programmers to grasp, especially when contrasted with
well-established notions such as linearizability used to reason about
shared-memory concurrency.  This state of affairs has made it
difficult to seamlessly adapt and exploit ongoing progress in the
development of scalable and correct concurrent algorithms used in the
shared-memory world to a geo-replicated setting.

In order to bridge this gap, we study how to \emph{automatically
  transplant} concurrent library implementations developed for shared
memory systems to replicated ones. Doing so would allow us to use
carefully-crafted implementations which have been proven to run
correctly in shared memory environments, thereby simplifying the task
of building distributed replication-aware applications.  However,
realizing this goal poses a number of challenges, the most critical of
which is the widely different memory consistency models used in the
two domains: the eventually consistent memory model typically provided
by a replicated system is significantly weaker than the sequential
consistency guarantees offered by shared-memory.  Consistency policies
offering session, causal, or transactional guarantees must be
additionally considered to facilitate correct behavior. This requires
enriching the \textit{semantics} of existing library implementations
to take into account the consistency policy of the underlying
replicated system.  Furthermore, the \emph{de facto} correctness
criterion for concurrent library implementations is linearizability,
which is clearly too restrictive to be directly applied to this much
weaker setting, since it demands that any correct execution be
equivalent to some sequential execution of a reference implementation.
Such a requirement is problematic in a geo-replicated environment
where the cost of coordination to enforce a global ordering of all
actions is prohibitive.  These observations are similar to those made
by Raad \emph{et al.}\cite{RDR19} who considered the applicability of
linearizability in a weak memory context, a scenario that faces
similar challenges to our own.  To address these issues, we therefore
consider alternative declarative specifications of data structures,
based on axiomatic definitions\cite{EEH15}, that are roughly
equivalent to the guarantees provided by linearizability (and hence
familiar to programmers), but suitably relaxed to take into account
the weak behaviors admitted by replicated systems.

We then propose an automated approach to find bounded violations of
these declarative specifications given an implementation and a
consistency policy.  Due to the non-deterministic nature of replicated
systems, manifesting violations in actual executions requires (1) a
specific combination of library methods to be called (2) with specific
argument values and (3) a specific interaction of low-level read/write
events.  Indeed, existing approaches to checking application safety
under weak consistency~\cite{Jepsen} potentially involve long (on the
order of hours) and costly execution runs to offer meaningful
assurance on application correctness given the large space of possible
behaviors that can be exhibited.



In contrast to testing approaches, our analysis framework
directly searches for an execution violating a specification, and in
the process \emph{constructs} the combination of library methods to be
called as well as their argument values, and the low-level read/writes
which can lead to the violation.  Moreover, because our analysis is
parametric in the choice of consistency policy, we can constrain the
search for violating executions on-demand as per the chosen policy.
We additionally show how our technique is capable of expressing
complex correctness specifications of libraries (see \S 3.4) and how
it can be used to automatically find violations in the face of this
complexity.  The analysis is sound in that it only reports actual
violations.  Notably, our experiments manifest a number of non-trivial
and complex violating executions for realistic concurrent libraries
which require intricate interaction with library methods.  We were
also able to analyse application behavior under different consistency
policies, and in particular, were able to find the weakest consistency
policy to eliminate a particular violation. Our analysis is based on
developing an efficient encoding of the implementation, the
consistency policy, and the correctness specification as first-order
logic formulae which can be dispatched to off-the-shelf SMT solvers to
find violations.  Unlike random testing approaches, our technique is
capable of identifying non-trivial subtle safety violations in the
order of minutes, making it feasible to use not only for finding
violations, but also for checking the feasibility of any proposed
remediations.

We make the following major contributions:

\begin{enumerate}
\item We propose a novel operational semantics for replicated systems
  operating parameterized under realistic consistency policies which
  can be used to describe executions of sophisticated concurrent
  library implementations.
  
\item We demonstrate how to adapt existing specification frameworks
  developed for concurrent libraries on shared memory systems to
  replicated systems with minimal changes.

\item We describe an automated bounded verification procedure to
  detect violations of such specifications for implementations
  intended to execute under a given consistency policy.

\item We catalog the results of applying our analysis on a number of
  well-studied implementations including stacks, queues and
  exchangers, on a commercial replicated store (Cassandra),
  demonstrating empirically that our correctness checking procedure is
  useful in practice.
  
\end{enumerate}

\noindent The remainder of the paper is organized as follows.  In the next
section, we provide a motivating example to illustrate the challenges
of reasoning about concurrent libraries in a weakly-consistent
replicated environment.  Section~\ref{sec:lang} formalizes the
language used to write library implementations and the specifications
that characterize their intended behavior.  Section~\ref{sec:verify}
describes our bounded verification procedure and provides details
about how we encode extracted verification conditions.
Section~\ref{sec:experiments} describes experimental results and
presents case studies to illustrate the effectiveness of our approach.
Related work and conclusions are given in Section~\ref{sec:related}.

\section{Illustrative Example}
\label{sec:example}

\begin{figure}
\vspace*{-.3in}
\begin{minipage}{.45 \textwidth}
\begin{lstlisting}
push(v){
1:  n = New(Node);
2:  n.Val = v;
    while(true){
3:    t = Top;
4:    n.Next = t;
5:    if (CAS(Top, t, n)) 
         break;
  }
}
\end{lstlisting}
\end{minipage}\hfill
\begin{minipage}{.45 \textwidth}
\begin{lstlisting}
pop(v){
  while(true){
6:    t = Top;
      if (t == NULL)
         return EMPTY;
7:    v = t.Val;
8:    n = t.Next;
9:    if (CAS(Top, t, n))
         return v;}
}
\end{lstlisting}
\end{minipage}
\caption{Treiber Stack}
\label{fig:impl}
\vspace*{-.2in}
\end{figure}

In this section, we illustrate the various issues that arise when
running standard concurrent library implementations on replicated
systems. Fig. \ref{fig:impl} shows the implementation of a {\sf
  Treiber stack}, suitably adapted to execute in a replicated
environment. The {\sf Treiber stack} provides two methods (\C{push}
and \C{pop}) to clients, and stores the elements of the stack in a
linked list, with the order of elements in the list corresponding to
the order in which elements are pushed. Since replicated stores
typically offer a database or a key-value store interface, we store
the linked list as a table of type \C{Node} with columns \C{Val} and
\C{Next}, where each row stores a node of the linked list, with
\C{Val} storing the value and \C{Next} storing the id of the next
node. \C{Top} contains the id of the \C{Node} row which is current top
of the stack (\C{Top} is initialized with the special value \C{NULL}
indicating an empty stack). In Fig. \ref{fig:impl}, variables denoted
by lower-case letters are assumed to be stored locally and are not
replicated. \C{New(Node)} returns the id of a new row in the \C{Node}
table. \C{CAS(Top, t, n)} is the typical Compare-And-Swap operation
which atomically compares \C{Top} to \C{t}, and if it is equal to
\C{t} then updates it to \C{n}\footnote{CAS operations are typically
  supported in replicated systems by providing transactional
  guarantees to a group of operations; e.g., lightweight transaction
  support provided in Cassandra\cite{LIGHT}.}.

\begin{wrapfigure}{r}{.6\textwidth}
\vspace*{-.3in}
\includegraphics[scale=.3]{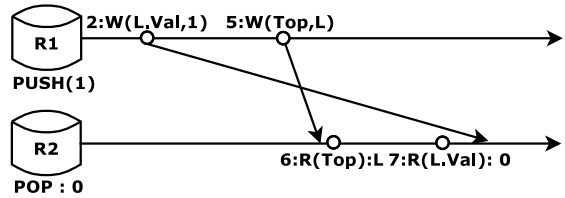} 
\caption{An Execution of Treiber Stack on a Replicated Store}
\label{fig:ex1}
\vspace*{-.3in}
\end{wrapfigure}

Clients of concurrent libraries issue invocations of a data
structure's methods, possibly at different replicas, with invocations
being grouped together into \emph{sessions}, with each session
containing invocations issued by the same client.  Whenever a method
is invoked, the underlying implementation of the method is executed;
we assume the various reads and writes performed by the method may
possibly be executed at different replicas.  All low-level operations
performed by the same invocation are defined to be in the same session
(i.e. the session of the parent invocation).  Notice that the
implementation stores data across a number of locations (e.g. \C{Top}
or a cell in the \C{Node} table), each of which are operated
independently through low-level read/write/CAS operations.  The replicated
store only guarantees eventual consistency, which means that the
values stored at all locations eventually converge across all
replicas.  However, users expect the behavior of the library to
conform to the specification of the stack data structure, regardless
of when and how updates propagate across replicas.

Consider the following basic specification (adapted from the
$\F{AddRem}$ axiom in \cite{EEH15}), which simply says that any value
returned by a $\F{POP}$ operation must have been pushed by some
$\F{PUSH}$ operation in the execution; observe that the specification
does not allude to any specific system-level issues related to replication
or weak consistency:
\[
\forall \gamma. \F{meth}(\gamma) = \C{POP}\ \wedge \F{ret}(\gamma) \neq \C{EMPTY}\ \Rightarrow
\exists \gamma'. \F{meth}(\gamma') = \C{PUSH}\ \wedge \F{arg}(\gamma') = \F{ret}(\gamma)
\]
Consider the execution shown in Fig \ref{fig:ex1} that involves an
invocation of $\F{PUSH}(1)$ and $\F{POP}$ from two different
replicas. Among the many operations that the implementation of
$\F{PUSH}$ performs, we show only two write operations in the figure
(along with line numbers referring to the implementation in
Fig.~\ref{fig:impl}), namely the write to the \C{Val} field of
location $\F{L}$ ($\F{L}$ is the id of the new \C{Node}), and the
write to \C{Top} as a result of the successful CAS. Similarly, for the
$\F{POP}$ operation, we show the read to \C{Top}, and then the read to
the \C{Val} field. In the execution, the write to \C{Top} propagates
from replica R1 to R2 before the read, but the write to \C{Val} does
not, so that $\F{POP}$ sees that a new node has been pushed but does
not read the value that was actually pushed, instead returning the
initial value of the location, thus breaking the specification
described above.  Eventual consistency only guarantees that
eventually, the write to \C{Val} will also be propagated to R2, which
is not sufficient to guarantee the specification holds under all
executions.

One way to avoid this counterexample would be to ensure that the write
to \C{Val} field by $\F{PUSH}$ is propagated to another replica before
the write to \C{Top}, thus guaranteeing that it would be available to
the read of \C{Val} by $\F{POP}$. Notice that the write to \C{Val}
occurs before the write to \C{Top} in the same session, and hence we
can use session guarantees to ensure the required behavior. In
particular, under a \emph{Monotonic Writes} (MW) consistency policy,
writes are always propagated in their session order to all
replicas~\cite{BDF13}. However, MW is not sufficient by itself to
eliminate the counterexample since the reads to \C{Top} and \C{Val} by
$\F{POP}$ may occur at different replicas, so that the read to \C{Val}
may occur at a replica in which none of the writes by $\F{PUSH}$ have
propagated. Hence, we also need to have these operations execute under
a \emph{Monotonic Reads} (MR) consistency policy that mandates all
writes witnessed by an operation will also be witnessed by later
operations in the same session.\footnote{We formalize all consistency
  policies used in the paper in the next section.}

\begin{figure}
\vspace*{-.2in}
$\xymatrix@R=5pt{
\underline{\F{PUSH}(1)} & \underline{\F{PUSH}(2)} & \underline{\F{POP}:2} & \underline{\F{POP}:0}\\
*+[F]{\C{2}:W(\C{L}_1.\C{Val}, 1)} & \C{3}:R(\C{Top}): \C{L}_1 & \C{6}:R(\C{Top}):\C{L}_2 & \C{6}:R(\C{Top}):\C{L}_1\\
\hspace*{-8pt}\C{5}:W(\C{Top}, \C{L}_1) \ar[ru] & \C{5}:W(\C{Top}, \C{L}_2) \ar[ru] & \hspace*{3pt}\C{9}:W(\C{Top}):\C{L}_1 \ar[ru] & *+[F]{\hspace*{7pt}\C{7}:R(\C{L}_1.\C{Val}):0}
}$
\caption{A Violation of $\F{AddRem}$ by Treiber Stack under MW+MR}
\label{fig:ex2}
\vspace*{-.2in}
\end{figure}
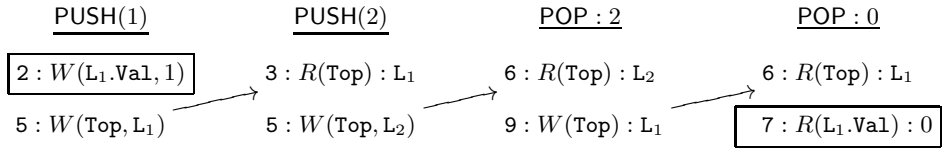

Hence, a combination of MW+MR prevents the counterexample in Fig.
\ref{fig:ex1}, but it is unfortunately not enough to guarantee the
$\F{AddRem}$ specification is correctly enforced.  Consider the
execution in Fig. \ref{fig:ex2} which involves four method invocations
(2 $\C{Push}$es and 2 \C{Pop}s), where each invocation occurs on a
different replica.  Again, we only show some relevant low-level
operations performed by these invocations, with arrows from write to
read operations showing reads-from ($\F{rf}$) dependencies. In the
execution, after the two pushes, 2 is stored on the top of stack at
\C{Node} $\C{L}_2$. Thus, the first \C{Pop} operation returns 2 and
sets the \C{Top} to point at $\C{L}_1$, which is then read by the
second \C{Pop}. However, MW+MR only guarantees that all write
operations performed by the first \C{Pop} will be witnessed by the
second \C{Pop}. Hence, just like in Fig.  \ref{fig:ex1}, the second
\C{Pop} operation may see the node at location $\C{L}_1$ but not the
write to the \C{Val} field (which was performed by \C{PUSH}(1)),
resulting in violation of the specification. To avoid this, it must be
guaranteed that the write to $\C{L}_1.\C{Val}$ by \C{Push(1)} must be
visible to its read by the second \C{Pop} (depicted by the two boxes
in Fig. \ref{fig:ex2}). This can be guaranteed by the \emph{Write
  Follows Read} (WFR) policy, which analogously to MW, ensures that
writes witnessed in a session are propagated to all replicas before
writes of the session itself (as opposed to MW which only ensures that
writes performed in a session are propagated in session order). We
note that both the violations described above (along with their
repairs) were automatically discovered using our proposed methods,
which devised solutions significantly less expensive than imposing
strong consistency (aka global coordination) on all accesses.

While MW+MR+WFR is required to ensure $\F{AddRem}$ in a {\sf Treiber
  Stack}, we found that weaker consistency policies (including
\emph{Eventual Consistency}) were sufficient for other properties and
benchmarks (more details are provided in \S 5).





\section{Semantics and Specifications}
\label{sec:lang}

In this section, we define a simple language to write library
implementations, powerful enough to express a number of real-world
implementations.  We then define an operational semantics to express
executions of any implementation written in the proposed language on
top of a replicated store. A key feature of this operational semantics
is that it is parametric in the consistency policies available to the
store. Thus, instantiating the semantics with different consistency
policy definitions allows us to reason about library behavior under
replicated stores providing different consistency guarantees. Another
important feature of the semantics is that it abstracts out low-level
operational details such as the number of replicas, the specific
manifestation of how message sends and receives are implemented,
etc., and instead uses a succinct representation involving read and
write events (and various binary relations among them) to capture
salient characteristics sufficient to reason about library correctness
with respect to consistency properties. The proposed semantics
facilitates a bounded verification approach
that is parametric in the consistency policy, and also matches very well
with existing axiomatic approaches to specify correctness of library implementations
in shared memory systems.


First, we define a simple imperative language in which implementations can be written:

\begin{mathpar}
\small
\begin{array}{lcl}
\multicolumn{3}{c}{
\C{v} \in \C{LocalVar} \qquad \C{l} \in \mathtt{Locations}   } \\
\multicolumn{3}{c}{
\oplus \in \{+, -, \times, /\} \qquad \odot \in \{<, \leq, ==, >, \geq\} \qquad \circ \in \{\wedge, \vee\}
}\\
e & \coloneqq & e \oplus e \ALT \C{v}\\
b & \coloneqq & b \circ b \ALT e \odot e\\
c & \coloneqq & \C{v} = e \ALT \C{v} = \C{l} \ALT \C{l} = e \ALT \C{If } b \C{ then } c \C{ else } c \\
  &   & \ALT c;c \ALT \C{while } b \C{ do } c \ALT v = \C{CAS}(\C{l},e_1, e_2)  \\
& &  \ALT \C{return } e \ALT \C{return}
\end{array}
\end{mathpar}

The only difference between standard shared-memory programs and those
written in the above language is that read and write operations can
now be performed on either \C{Locations}, which are replicated, or
local variables which are not. As we saw in \S 2, replicated
\C{Locations} can in general refer to any field in any table. Let
$\mathbb{P}$ be the set of programs ($c$) generated using the above
grammar. A \textbf{library} $L = (M, I)$ consists of a set of methods
($M$) and an implementation function $I : M \rightarrow
\mathbb{P}$. For simplicity, we assume that each method takes as input
one argument. Assume that $I(m)$ contains the free variable $\C{a}$
that stores the input argument. Let $\mathbb{V}$ be the value
domain for arguments and return values. We designate a special value
$\bot \in \mathbb{V}$ for the cases where the argument or return value
is empty.

The methods of a library implementation $L$ can be invoked any number
of times by multiple clients. Invocations from the same client are
grouped together into \textbf{sessions}, where each session consists
of a sequence of method invocations. Following standard terminology,
given a set of sessions $S$, an interaction between clients and the
library is expressed as a \textbf{history}, $h: S \rightarrow (M
\times \mathbb{V})^*$, which simply associates a sequence of methods
invocations to each session. An execution of the history corresponds
to executing the library implementation of each method in the history
on the replicated store. The store constrains the behavior of reads,
writes and CAS operations to replicated \C{Locations} through its
consistency policy.

We now formally define the operational semantics of a history on a
replicated store that is parametric in a consistency policy
$\Psi$. While the history only associates arguments with method
invocations, executing it on the replicated store will give rise to an
\textbf{abstract execution}, which will also associate return values
with invocations, and whose correctness we are interested in
checking. Given a history $h$, library $L$, and consistency policy
$\Psi$, we define our semantics in terms of a labeled transition system (LTS)
$\Omega_{h,L,\Psi} = (\Phi, \mathcal{E}, \rightarrow)$, where $\Phi$
denotes a set of states, $\mathcal{E}$ denotes a set of events (also
used as labels) and $\rightarrow \subseteq \Phi \times \mathcal{E}
\times \Phi$ defines a transition relation over states and events.

Each state in $\Phi$ is specified as a tuple $(\chi, h', \mu, \C{c}, \alpha)
$. $\chi$ denotes the replicated store state and consists of
read/write/update events to \C{Locations} and various relations among
them (described in detail later); $h' : S \rightarrow (M \times
\mathbb{V})^*$ denotes the continuation of the history, i.e., the
remaining history yet to be executed; $\mu : S \rightarrow (\C{LocalVar} \rightarrow \mathbb{V})$
denotes the local variables map for each session;
$\C{c} : S \rightarrow \mathbb{P}$ denotes the
continuation of the current invocation for each session, i.e., the
implementation of the current invocation for each session that is yet
to be executed and $\alpha$ denotes the abstract
execution. Each \textbf{event} $\sigma \in \mathcal{E}$ is a
tuple $(i,s,a)$, where $i$ is a unique event-id, $s \in S$ is the
session from which the event originated, and $a$ is the action to the
replicated store (either read $\F{R}(l,n)$, write $\F{W}(l,n)$ or
update $\F{U}(l,m,n)$). Given an event $\sigma=(i,s,a)$, $act(\sigma)$
denotes the action $a$, $loc(\sigma)$ denotes the location that is the
subject of the action.

\subsection{Language Semantics}

To simplify the presentation, we decouple the semantics of the
language from the semantics of the replicated store. The language is 
defined via a standard imperative semantics \emph{except}
that there are no constraints on reads to replicated locations (i.e.,
we do not mandate a specific replica that is targeted by the read),
and every operation to a replicated location generates an event.
These rules do not concern the replicated store state, and hence are
of the form $(h_1, \mu_1, c_1, \alpha_1) \xrightarrow{\sigma} (h_2,
\mu_2, c_2, \alpha_2)$ (i.e. omitting $\chi$ from $\Phi$). We essentially
pick any session and then execute the next operation from the current
invocation in the session, or initiate the next invocation in the
session if there is no invocation currently running. As an
illustration, consider the following rule \rulelabel{L-Read}: \\
$$
\RULE{\C{c}(s) \equiv \C{v}=\C{l};\C{c}' \quad \sigma = (\F{i},s,\F{R}(l,n)) \quad \F{fresh}\ \F{i} }{(h',\mu,\C{c},\alpha) \xrightarrow{\sigma} (h', \mu[s \rightarrow \mu(s)[\C{v} \rightarrow n]], \C{c}[s \rightarrow \C{c}'],\alpha)}
$$

The rule picks the next operation in session $\F{s}$ which is a read
operation to location $\C{l}$, and generates the read event $\sigma$
reading value $\F{n}$ from $\C{l}$. It updates the local variable
$\C{v}$ to this value, 
leaving the yet-to-be-executed history ($h'$)
and abstract execution ($\alpha$) unchanged. Write statements
(i.e. $\C{l} = n$) generate write events ($\F{W}(l,n)$), successful
CAS statements (i.e. $\C{v} = \C{CAS}(\C{l}, m, n$) generate update
events ($\F{U}(l,m,n)$), and unsuccessful CAS generates read events
($\F{R}(l,m')$). The complete set of rules can be found in the
Appendix A.1.

\subsection{Abstract Execution Semantics}

An \textbf{abstract execution} $\alpha = (\Gamma, \F{so}_{\Gamma})$
maintains a set of method invocation events in $\Gamma$ and a session
order relation $\F{so}_{\Gamma}$ among these events. Each method
invocation event $\gamma \in \Gamma$ is a tuple $(i,m,a,r,s)$ where
$i$ is a unique event-id, $m \in M$ is a method of the library, $a,r
\in \mathbb{V}$ are the method argument and return values respectively
and $s \in S$ is the session from which the method was called. We use
the notation $\Gamma^{s}$ for the subset of $\Gamma$ which only
contains method invocation events that originate in session $s$. The
following rule (\rulelabel{L-Return-Val}) describes the generation of
a method invocation event, which occurs on encountering a \C{return}
statement during execution, and which is added to the abstract
execution.
$$\RULE{\C{c}(s) \equiv \C{return } e;c' \quad h'(s) = m(k) \cdot h'' \quad \llbracket e \rrbracket_{\mu(s)} = n \\
\alpha = (\Gamma, \F{so}_{\Gamma}) \quad \gamma = (i,m,k,n,s) \quad \alpha' = (\Gamma \cup \{\gamma\}, \F{so}_{\Gamma} \cup \Gamma^{s} \times \{\gamma\} )}{(h', \mu, \C{c},\alpha) \rightarrow (h'[s \rightarrow h''], \mu, \C{c}[s \rightarrow \epsilon],\alpha')}$$

The rule updates the yet-to-be executed history $h'$ by removing the
current invocation $m(k)$ (since this invocation has now completed),
updates the abstract execution $\alpha$ to now include the newly
completed invocation, and updates the current invocation
implementation to empty. Note that $\llbracket e \rrbracket_{\mu(s)}$
denotes the evaluation of the expression $e$ under the local variable map $\mu(s)$.
 When the history $h'$ becomes empty,
i.e. there are no more method invocations to be executed, the abstract
execution becomes complete and would include all method instances
present in the original history $h$. Note that this rule does not
generate any read/write/update event.

\subsection{Replicated Store Semantics}

The replicated store state $\chi = (\Sigma, \F{vis}, \F{ar}, \F{so})$
consists of the set of replicated store events ($\Sigma$) and various
relations on $\Sigma$. Events can either be read, write or update
events, and depending on the type of event, $\Sigma$ is partitioned
into $\Sigma_R, \Sigma_W$ and $\Sigma_U$. The visibility relation
$\F{vis} \subseteq \Sigma \times \Sigma$ denotes
the events visible to an event and is used to determine the output of read events. 
The arbitration
relation $\F{ar} \subseteq (\Sigma_W \cup \Sigma_U) \times (\Sigma_W
\cup \Sigma_U)$ provides a total ordering on write or update events to
the same location. Finally, the session order relation $\F{so}
\subseteq \Sigma \times \Sigma$ provides a total ordering on events
originating from the same session. All events generated by statements
in the same method invocation would belong to the same session and
hence would be related by $\F{so}$. We also define a happens-before
relation $\F{hb} = (\F{vis} \cup \F{so})^{+}$ in the usual way.

We use $\Psi$ to refer to a consistency policy supported by the
store. $\Psi$ is a predicate on the store state, which must be
maintained at every step of the execution. $\Psi$ essentially controls
the visibility relation on events based on session or happens-before
order. The following table illustrates the various consistency
policies that we consider in our work; all of these policies can be
implementation without the need for global
coordination~\cite{BDF13}.\footnote{Note that the lack of any
  constraints (i.e. $\Psi = true$) corresponds to Strong Eventual
  Consistency~\cite{GO17}. Since we assume SEC, our definition of Causal Consistency corresponds to Causal Convergence (\C{CCv}) as defined by \cite{BEGH17}} (all $\sigma_i$ belong to $\Sigma$):

\begin{table}[h]
\vspace*{-.2in}
\begin{center}
\small
\begin{tabular}{| l | c |}
\hline
\textbf{Consistency Policy} & $\bm{\Psi(\Sigma, \F{vis}, \F{ar}, \F{so})}$\\
\hline \hline
{\sf Read Your Writes} \cite{TDP94} & $\F{so}(\sigma_1, \sigma_2) \Rightarrow \F{vis}(\sigma_1, \sigma_2)$\\ \hline
{\sf Monotonic Writes} \cite{TDP94} & $\F{so}(\sigma_1, \sigma_2) \wedge \F{vis}(\sigma_2, \sigma_3) \Rightarrow \F{vis}(\sigma_1, \sigma_3)$ \\ \hline
{\sf Monotonic Reads} \cite{TDP94} & $\F{vis}(\sigma_1, \sigma_2) \wedge \F{so}(\sigma_2, \sigma_3) \Rightarrow \F{vis}(\sigma_1, \sigma_3)$ \\  \hline
{\sf Write Follow Read} \cite{TDP94} & $\F{vis}(\sigma_1, \sigma_2) \wedge \F{so}(\sigma_2, \sigma_3) \wedge \F{vis}(\sigma_3, \sigma_4) \Rightarrow \F{vis}(\sigma_1, \sigma_4)$ \\ \hline
{\sf Causal Visibility} \cite{LL11} & $\F{hb}(\sigma_1, \sigma_2) \wedge \F{vis}(\sigma_2, \sigma_3) \Rightarrow \F{vis}(\sigma_1, \sigma_3)$ \\ \hline
{\sf Causal Consistency} \cite{LL11} & $\F{hb}(\sigma_1, \sigma_2) \Rightarrow \F{vis}(\sigma_1, \sigma_2)$\\ \hline
\end{tabular}
\end{center}
\caption{Axiomatic characterization of various weak consistency policies.}
\label{tab:consistency}
\vspace*{-15pt}
\end{table}

As we saw earlier in \S 2, ${\sf Monotonic Writes}$ enforces the
constraint that if an event is visible, then all events before it in
session order must also be visible. ${\sf Monotonic Reads}$ requires
that if an event is visible, it will continue to remain visible to all
operations later in the session. On the other hand, ${\sf Write
  Follows Read}$ enforces that all events visible to a prior event in
a session will continue to remain visible to other events which
witness a later event of the session.

We use the notation $\Sigma^{l}$ to denote the subset of events
pertaining to location $l$, and $\Sigma^{s}$ to denote the subset of
events of session $s$. Given a set of events $\Sigma'$,
$\F{MAX}^{\C{l}}_{\F{ar}}(\Sigma')$ denotes the maximal events in
$\Sigma'$ according to the relation $\F{ar}$ which write to location
\C{l}. Given events $\sigma \in \Sigma_R^l$, $\sigma' \in \Sigma_W^l$, we define the
\emph{Reads-From} relation $\F{rf}$ in terms of $\F{vis}$ and $\F{ar}$
relations as follows:
\begin{mathpar}
\begin{array}{lcl}
\F{rf}(\sigma',\sigma) & \Leftrightarrow & \F{vis}(\sigma', \sigma) \wedge \forall \sigma'' \in \Sigma^l. (\F{vis}(\sigma'',\sigma) \wedge \sigma'' \neq \sigma ) \Rightarrow \F{ar}(\sigma'', \sigma'))
\end{array}
\end{mathpar}

The $\F{rf}$ relation essentially encodes the `last writer wins'
nature of the store, whereby the most recent visible write event
according to $\F{ar}$ becomes the event supplying the value available
to subsequent reads. The replicated store state evolves by the
addition of new events. On addition of a write/update event,
the arbitration order is appropriately modified to ensure that 
it remains a total order on events targeting the same location. In addition,
we also ensure causal arbitration \cite{B14} by enforcing that $\F{ar}$ and $\F{hb}$
do not disagree with each other. For update and read events, 
the values that these events read depend upon the most recent write
event to the same location visible to the events, which in turn is
controlled by the consistency policy. To elaborate, consider the rule
\rulelabel{R-CAS}:
$$
\RULE{\Sigma' \subseteq \Sigma \quad \sigma' \in \F{MAX}^{l}_{\F{ar}}(\Sigma') \quad \F{ar} \subseteq \F{ar}' \\
act(\sigma') = \F{W}(l,m) \vee act(\sigma') = \F{U}(l,\_,m) \quad \sigma = (i,s,U(l,m,n)) \quad \forall \tau \in \Sigma^{l}_{U}. \neg(\F{rf}(\sigma', \tau)) \\
  \F{ar}'\ \F{is}\ \F{a}\ \F{total}\ \F{order}\ \F{on}\ \Sigma^l \cup \{\sigma\} \quad \forall \sigma_1,\sigma_2. \neg(\F{hb}(\sigma_1, \sigma_2) \wedge \F{ar}'(\sigma_2, \sigma_1))  \\
vis' = vis \cup \Sigma' \times \{\sigma\} \quad so' = so \cup \Sigma^s \times \{\sigma\} \quad \Psi(\Sigma \cup \{\sigma\}, \F{vis}', \F{ar}', \F{so}')}{
(\Sigma, \F{vis}, \F{ar}, \F{so}) \xrightarrow{\sigma} (\Sigma \cup \{\sigma\},\F{vis}',\F{ar}',\F{so}'  )}
$$

Here, we want to add a new update event to location $l$. First, an
\textit{arbitrary} subset ($\Sigma'$) of events of
$\Sigma$ is selected. This step essentially corresponds to the creation of a new replica
on which the events in $\Sigma'$ have been applied.
 Then, we select the most recent write event
($\sigma'$) from $\Sigma'$ which ensures atomicity of the update event
(and hence the CAS statement responsible for the update). In
particular, we require that no other update event must have read from
($\F{rf}$) $\sigma'$. The value written by $\sigma'$ (i.e. $m$) would
be the read value of the update event. $\F{vis}$, $\F{so}$ and
$\F{ar}$ are appropriately updated, and the new store state must
satisfy the consistency policy $\Psi$, which in turn will govern the
selection of the initial subset $\Sigma'$. The formal rules for read
and write events can be found in Appendix A.2. 

Note that enforcing the above rule would in essence prohibit two CAS
operations to be executed concurrently, and hence would establish a
global ordering among the CAS operations. However, unlike in shared
memory systems where this is sufficient to establish a global ordering
among all operations thus ensuring linearizability, in replicated
systems, this does not constrain the behavior of other read and write
operations (as we saw in \S 2), and hence more constraints must be
enforced through the consistency policy.

We can now combine the language, abstract execution, and replicated
store rules to describe transitions of the LTS $\Omega_{h,L,\Psi}$,
which simply requires the language rules and the replicated store
rules to agree on the structure of all replicated store events:
$$
\RULE{(h', \mu, \C{c}, \alpha) \xrightarrow{\sigma} (h'', \mu', \C{c}', \alpha) \quad \chi \xrightarrow{\sigma} \chi'}{(\chi, h', \mu, \C{c}, \alpha) \xrightarrow{\sigma} (\chi', h'', \mu', \C{c}', \alpha)}
$$

$$
\RULE{(h', \mu, \C{c}, \alpha) \rightarrow (h'', \mu', \C{c}', \alpha')}{(\chi, h', \mu, \C{c}, \alpha) \rightarrow (\chi, h'', \mu', \C{c}', \alpha')}
$$

\textbf{Example:} Let us revisit the Treiber Stack and in particular
the violating execution described in Fig. \ref{fig:ex1}. The violating
history consists of two sessions, with one session containing the
invocation \C{push(1)} and another containing \C{pop}. The execution
of \C{push(1)}, following the language semantics, creates the events
$\sigma_1$ and $\sigma_2$ such that $act(\sigma_1) =
\F{W}(L.\C{Val},1)$ and $act(\sigma_2) = \F{U}(\C{Top}, \C{NULL},
\F{L})$ which are both added to the store state. The execution of
\C{pop} generates the read event to \C{Top}, which following the store
semantics picks the set $\Sigma' = \{\sigma_2\}$, resulting in read
event $\sigma_3$ such that $act(\sigma_3) = \F{R}(\C{Top},
\F{L})$. Under {\sf EC}, the following read to L.\C{Val} by \C{pop} is
unconstrained and hence simply picks $\Sigma' = \phi$, resulting in
the event $\sigma_4$ such that $act(\sigma_4) = \F{R}(L.\C{Val}, 0)$
where 0 is the initial value. This results in violation of the {\sf
  AddRem} specification.

Notice that $\F{so}(\sigma_1, \sigma_2)$ and $\F{vis}(\sigma_2,
\sigma_3)$. Hence, under {\sf MW+MR}, while generating the read event
to L.\C{Val} by \C{pop}, the store must pick $\Sigma' = \{\sigma_1,
\sigma_2\}$ to satisfy the axioms of {\sf MW+MR}, so that the event
must read the value 1, which prevents the violation from occurring.

\subsection{Correctness Specification}

Given an abstract execution obtained after executing a history on a
replicated store under some consistency policy, how do we decide if it
correctly obeys the semantics of the data structure implemented by the
library? Linearization would require us to demonstrate a total order
on all method invocations which would be admissible by a sequential
reference implementation of the data structure. However, since the
consistency model of a replicated system is substantially weaker than
sequential consistency, it becomes necessary to also weaken
correctness requirements \cite{RDR19,S11}. We use the axiomatic
specifications of data structure correctness as proposed by Emmi
et. al.\cite{EEH15}, which are equivalent to standard linearizability,
as our basis, and then weaken them systematically to adapt them to be
useful in a replicated environment. Axiomatic specifications do not
require a total order to be established on method invocations, do not
refer back to a reference implementation, and also match the
axiomatic, declarative nature of the semantics of the replicated
store.

First, we define all abstract executions that can be generated given a
library implementation, a history and a consistency policy. The
initial state of the replicated store is assumed to be empty,
i.e. $\chi_{\F{Init}} = (\phi, \phi, \phi, \phi)$. Let $h_{\epsilon}$
be the empty history which associates an empty sequence ($\epsilon$)
of invocations to each session. Let $\C{c}_{\F{Init}}$ be the initial
implementation state which simply associates the empty program
$\epsilon$ to each session.

\begin{definition}
Given a set of sessions $S$, a history $h$, a library implementation $L$ and a consistency policy $\Psi$, the abstract executions generated by $\Omega_{h, L, \Psi}$ are defined as : $\llbracket \Omega_{h,L,\Psi} \rrbracket = \{\Gamma\, |\, (\chi_{\F{Init}}, h, (\phi, \phi), \C{c}_{\F{Init}}) \rightarrow^{*} (\_, h_{\epsilon}, \Gamma, \_)\}$
\end{definition}

Thus, executing all invocations in the history under a given
consistency policy and library implementation gives rise to the set of
final abstract executions. Due to the non-deterministic nature of the
semantics, multiple abstract executions could be
generated. Correctness of an abstract execution is specified in terms
of various \textbf{axioms} that it must obey. An implementation is
correct under a consistency policy if for all possible histories, all
final abstract executions generated by the implementation obey the
axioms.

To illustrate, let us consider the {\sf Stack} data structure. It has
two methods $M=\{\C{Push}, \C{Pop}\}$. Given a method invocation event
$\gamma = (i,m,a,r,s)$, we assume projection functions for all the
respective components (e.g., \C{m}, \C{a}, and \C{r}). Further, we
assume a $\F{match}$ predicate relating two method invocation events
defined thus:
\[\F{match}(\gamma_1, \gamma_2) \Leftrightarrow \F{m}(\gamma_1) = \C{Push} \wedge \F{m}(\gamma_2) = \C{Pop} \wedge \F{a}(\gamma_1) = \F{r}(\gamma_2)\]

\noindent Let $\C{EMPTY}$ denote a special value signifying the empty
return value (see, e.g. the Treiber Stack impl. in Fig
\ref{fig:impl}). Consider an abstract execution $\alpha = (\Gamma,
\F{so}_{\Gamma})$.  We define the happens-before relation for method
invocations as $\F{hb}_{\Gamma} = (\F{match} \cup
\F{so}_{\Gamma})^{+}$. Then, the correctness of $\alpha$ can be
specified in terms of the following axioms:
\begin{itemize}
\item $\F{AddRem}$ : $\forall \gamma \in \Gamma. \F{m}(\gamma) = \C{Pop} \wedge \F{r}(\gamma) \neq \C{EMPTY} \Rightarrow \exists \gamma' \in \Gamma. \F{match}(\gamma', \gamma)$
\item $\F{Injective}$ : $\forall \gamma_1, \gamma_2, \gamma_3 \in \Gamma. \F{match}(\gamma_1, \gamma_2) \wedge \F{match}(\gamma_1, \gamma_3) \Rightarrow \gamma_2 = \gamma_3$
\item $\F{Empty}$ : $\forall \gamma_1, \gamma_2, \gamma_3 \in \Gamma. \F{m}(\gamma_1) = \C{Pop} \wedge \F{r}(\gamma_1) = \C{EMPTY} \wedge \F{m}(\gamma_2) = \C{Push} \wedge \F{hb}_{\Gamma}(\gamma_2, \gamma_1) \Rightarrow \exists \gamma_3 \in \Gamma. \F{match}(\gamma_2, \gamma_3)$
\item $\F{LIFO-1}$ : $\forall \gamma_1, \gamma_2, \gamma_3 \in \Gamma. \F{m}(\gamma_1) = \C{Push} \wedge \F{match}(\gamma_2, \gamma_3) \wedge \F{hb}(\gamma_2, \gamma_1) \wedge \F{hb}(\gamma_1, \gamma_3) \Rightarrow \exists \gamma_4 \in \Gamma. \F{match}(\gamma_1, \gamma_4)$
\item $\F{LIFO-2}$ : $\forall \gamma_1, \gamma_2, \gamma_3, \gamma_4 \in \Gamma. \neg (\F{match}(\gamma_1, \gamma_4) \wedge \F{match}(\gamma_2, \gamma_3) \wedge \F{hb}(\gamma_2, \gamma_1) \wedge \F{hb}(\gamma_3, \gamma_4) \wedge \F{hb}(\gamma_1, \gamma_3))$
\end{itemize}
These axioms follow from those given in \cite{EEH15}, except that
instead of using a linearization order as done in~\cite{EEH15}, we use
a weaker happens-before $\F{hb}_{\Gamma}$ order. It is also possible
to use the even weaker session order $\F{so}_{\Gamma}$ in place of
$\F{hb}_{\Gamma}$. We have already seen the $\F{AddRem}$ axiom in \S
2. The $\F{Injective}$ axiom enforces that an element pushed onto the
stack is not popped more than once\footnote{Note that we assume 
all methods are called with distinct arguments}. The $\F{Empty}$ axiom says that if
a pop invocation ($\gamma_1$) returns \C{EMPTY} and if there is a push
invocation ($\gamma_2$) that happens-before it, then $\gamma_2$ must
be matched to another pop. This reflects the expected stack-like
behavior from the point of view of a client who observes these
invocations. The $\F{LIFO-1}$ property specifies that if a push
invocation $\gamma_2$ happens-before another push invocation
$\gamma_1$, with both of them happening-before a pop invocation
$\gamma_3$, and if $\gamma_2$ is matched with $\gamma_3$, then to
respect the LIFO order, $\gamma_1$ must also be matched (to some
$\gamma_4$). $\F{LIFO-2}$ complements $\F{LIFO-1}$ by requiring that
$\gamma_3$ cannot happen-before such a $\gamma_4$. The specifications
for other data structures we have considered, including {\sf Queue}
and {\sf Exchanger} can be found in Appendix A.3.

\section{Bounded Verification}
\label{sec:verify}

We now present an automated bounded verification procedure capable of
generating abstract executions that violate data structure correctness
specifications under a given consistency policy.  We take advantage of
the axiomatic nature of both the semantics and specification and
reduce the problem to that of checking the satisfiability of a
collection of formulae in first-order logic (FOL), which can be
dispatched to an off-the-shelf SMT solver. In particular, our strategy
is to instantiate a bounded number of invocations ($k$) without
specifying their method types, arguments, or session information, and
instead leave it upto the solver to search efficiently among all
histories of length $k$.
\subsection{Vocabulary}

Given a library $L = (\F{M}, \mathsf{Impl})$, we first take each
method implementation and unroll loops upto a constant
bound\footnote{Loops are typically only used to busy wait for a
  successful {\sf CAS} operation in the applications we consider.},
and give a label to each program statement that interacts with a
replicated location (e.g. see the Treiber Stack impl. in Fig
\ref{fig:impl}). Let $\mathbb{L}$ denote this set of labels.

We use an uninterpreted, finite sort $\invo$ to represent invocations
in the history that we wish to construct, and then constrain this sort
to contain only the distinct elements $\mathsf{INV}_1, \ldots,
\mathsf{INV}_k$. In addition, we use uninterpreted sorts $\event$ and
$\val$ to represent the set of replicated store events and values that
are read or written by them. We define the function $\mathsf{meth} :
\invo \rightarrow \F{M}$ to associate a method type with each
invocation. We use an uninterpreted sort $\sess$ to denote the set of
sessions involved in the history. The function $\mathsf{sess} : \invo
\rightarrow \sess$ associates a session with each invocation.

For each method $m \in \F{M}$ and each program statement labeled $n$
in the implementation $\mathsf{Impl}(m)$, we define the function
$\mathsf{P}_{mn}:\invo \rightarrow \event$ to associates the event
generated by the program statement to an invocation. In addition,
functions $\F{arg},\F{ret} : \invo \rightarrow \val$ associate the
argument and return values to each invocation. For every local
variable \C{v} used in a program, function $\rho_{\C{v}} : \invo
\rightarrow \val$ denotes the value of the local variable in that
invocation. The predicate $\F{so}_{\invo} : \invo \times \invo
\rightarrow \mathbb{B}$ denotes the session order relation among
invocation instances.

We define functions $\F{loc},\F{rval},\F{wval} : \event \rightarrow
\val$ to associate locations, values read and values written by events
resp. We use the uninterpreted, finite sort $\mathbb{E}$ containing
elements $\F{R}, \F{W}, \F{U}$ to denote various event types. The
function $\F{Etype} : \event \rightarrow \mathbb{E}$ associates the
type with each event. Finally, predicates $\F{vis}, \F{ar},
\F{so}_{\event}, \F{rf} : \event \times \event \rightarrow \mathbb{B}$
denote the visibility, arbitration, session order, and read-from
relations resp. among events.

For every replicated location, we also instantiate a distinct value
referring to the location. For example, for the Treiber Stack
implementation (Fig. \ref{fig:impl}), we have distinct values for
\C{Top} and for the \C{Val} and \C{Next} fields of each \C{New}
\C{Node} generated by an invocation. Since the number of invocations
is fixed ($k$), the number of such locations to be instantiated can
also be pre-determined statically. We also define a function
$\F{Initval} : \val \rightarrow \val$ which fixes an initial value for
every location, and assigns initial values to all locations used in
the execution.

\subsection{Implementation Constraints}

We now describe constraints on the events imposed by the
implementation. First, note that even though the set of functions
$\{\F{P}_{mn} | m \in \mathbb{M},\ n \in \mathbb{L}\}$ are defined for
every invocation, an invocation $\F{i}$ will only have a fixed method
type $\F{meth}(\F{i})$, and hence will only generate events
corresponding to program statements in the implementation of
$\F{meth}(\F{i})$. We designate a special event $\bot : \event$ and
associate it for program statements of every other method type using
the following constraint:

$$
\forall i \in \invo\ \forall m \in \mathbb{M}\ \forall n \in \mathbb{L}.\ m \neq \F{meth}(i) \Rightarrow \F{P}_{mn}(i) = \bot 
$$

\noindent For program statements in the implementation of $\F{meth}(\F{i})$, we
add constraints for every statement based on its type. Note that loops
have already been unrolled and for every statement labeled $n$ in
method $\F{m}$, we collect the conditionals of any \C{if} statement
enclosing the statement and replace any local variable $\C{v}$ used in
those conditionals with the corresponding function $\rho_{\C{v}}(\F{i})$
(for invocation $\F{i}$) to obtain the formulae $\llbracket \phi_{mn}
\rrbracket_{\F{i}}$. To illustrate the constraints added for different
types of statements, consider the rule for reads:

$$
\RULE{\F{Impl}(m):n:\ \C{v} = \C{l}}{\forall i \in \invo.\ (\F{meth}(i) = m \wedge \llbracket \phi_{mn} \rrbracket_{i}) \Rightarrow (\F{Etype}(\F{P}_{mn}(i)) = \F{R} \wedge  \F{loc}(\F{P}_{mn}(i)) = \C{l} \\ \wedge \F{rval}(\F{P}_{mn}(i)) = \rho_{\C{v}}(i)})
$$
 
The rule essentially specifies the constraint for statement labeled
$\F{n}$ in the implementation of method $\F{m}$ if it is a read
operation. The constraint appropriately sets the $\F{Etype}$,
$\F{loc}$ and $\F{rval}$ functions of event $P_{mn}(i)$ for every
invocation $i$, if the invocation has a method type of $\F{m}$ and the
enclosing \C{if} conditionals (if any) are satisfied. The rules for
write and CAS statements are similar (they also set the $\F{wval}$
function and additionally CAS also checks whether the value read is
equal to its first argument) and can be found in the Appendix B. In
addition, we also relate adjacent events of the same invocation with
the session order relation $\F{so}_{\event}$.
  
\subsection{Abstract Execution Constraints}

On encountering a \C{return} statement, we record the returned value using the following constraint:
$$
\RULE{\F{Impl}(m):n:\ \C{return } \C{v}}{\forall i \in \invo.\ (\F{meth}(i) = m \wedge \llbracket \phi_{mn} \rrbracket_{i}) \Rightarrow (\F{ret}(i) = \rho_{\C{v}}(i) \wedge \F{completed}(i))}
$$

Apart from setting the $\F{ret}$ value, we also use another unary
predicate $\F{completed}$ to encode that the invocation has completed
and reached the \C{return} statement. This is needed because we are
unrolling loops upto a fixed bound. Since we know the last program
statement statically, if we encounter this statement without reaching
\C{return} for an invocation, then $\F{completed}$ will be set to
$\F{false}$.

We also encode the constraint that the session order relation
($\F{so}_{\invo}$) among invocations of the same session is a total
order. Finally, we also encode that if two invocations $\F{i}_1$ and
$\F{i}_2$ are in session order ($\F{so}_{\invo}(\F{i}_1, \F{i}_2)$),
then the last event of $\F{i}_1$ and the first event of $\F{i}_2$ are
in event session order ($\F{so}_{\event}$).

\subsection{Replicated Store Constraints}

We must also encode constraints ensuring that the semantics of the
replicated store are preserved.  First, we capture various properties
of relations on events, viz. $\F{vis}$ is anti-symmetric and
irreflexive, $\F{ar}$ among write events to the same location is a
total order, $\F{vis}$ and $\F{so}_{\invo}$ do not clash with each
other, $\F{ar}$ does not clash with $\F{vis}$ and $\F{so}_{\invo}$. All
these constraints are implicitly enforced by the semantics of the
replicated store, so that the state of the store reached after any
number of execution steps must obey them.

The various consistency policies in Table \ref{tab:consistency} can be
directly encoded using the relations defined in the vocabulary. We now
turn to encoding the last-writer-wins nature of the data store, which
relates the $\F{vis}$ and $\F{ar}$ relations with the read and write
values ($\F{rval}$ and $\F{wval}$) of the events.
\begin{mathpar}
\begin{array}{lcl}
\forall e_1,e_2 \in \event. \F{rf}(e_1, e_2) & \Rightarrow & \F{vis}(e_1, e_2) \wedge \F{wval}(e_1) = \F{rval}(e_2) \wedge \\
& & \forall e_3 \in \event_{\F{W}}^{\F{loc}(e_2)}. (\F{vis}(e_3, e_2) \Rightarrow e_3 = e_1 \vee \F{ar}(e_3,e_1))
\end{array}
\end{mathpar}
$$
\forall e_1 \in \event_{\F{R}}. (\forall e_2 \in \event. \neg \F{rf}(e_2, e_1)) \Rightarrow \F{rval}(e_1) = \F{Initval}(\F{loc}(e_1))
$$ In the above constraints, we use the notation
$\event_{\F{W}}^{\F{l}}$ to indicate only those events that write to
location $\F{l}$, and $\event_{\F{R}}$ for read events. The first
constraint enforces the reads-from event to be the most recent visible
event according to the arbitration order, and also constrains the read
value. The second constraint disallows out-of-thin-air reads by
enforcing that if there are no $\F{rf}$ events, then the value read
must be the initial value. As an optimization, while encoding this
constraint in our tool, we enumerate all possible write events to the
same location (which are guaranteed to be finite since we only have
$k$ invocations) in the antecedent, instead of the universal
quantification used above.

For CAS operations which generate update events, we encode the constraint (as derived from the semantics rule R-CAS) that two update events should not read from the same event:
\[\forall e,e_1,e_2 \in \event.\ \F{Etype}(e_1) = \F{U} \wedge \F{Etype}(e_2) = \F{U} \wedge \F{rf}(e,e_1) \wedge \F{rf}(e,e_2) \Rightarrow e_1 = e_2
\]

\subsection{Specification Constraints}

The axioms of correctness for data structures only use an invocation's
argument and return values, and the session order relation among
invocations in the abstract execution.  Thus, they can be directly
encoded using our vocabulary. Given an axion $\theta$, we encode its negation to
find histories which have abstract executions that violate the axiom.

For example, to find violations of the $\F{AddRem}$ axiom, we add the following constraint:
\[
\exists \F{i}_1 \in \invo.\ \F{meth}(\F{i}_1) = \C{POP} \wedge \F{ret}(\F{i}_1) \neq \C{EMPTY} \wedge \forall \F{i}_2 \in \invo.\ \neg \F{match}(\F{i}_2, \F{i}_1)
\]
where we use the predicate $\F{match}: \invo \times \invo \rightarrow
\mathbb{B}$ defined in a similar manner as in \S 3.4. This completes
the entire description of our encoding.

Our main soundness result can be formalized thus\footnote{A Proof Sketch can be found in Appendix C}

\begin{theorem}
Given a library implementation $L$, consistency policy $\Psi$ and a
correctness axiom $\theta$, if the collection of formulae described
above are satisfiable, then there exists a history $h$ and an abstract
execution $\Gamma \in \llbracket \Omega_{h,L,\Psi} \rrbracket$ which
violates $\theta$.
\end{theorem}

\section{Experimental Evaluation}
\label{sec:experiments}


\begin{table}
  \tiny
  \centering
    \vspace*{-.2in}
  \begin{tabular}{| c | c | c | c | c | c | c | c |}
    \hline
{\bf Benchmark} & {\sf AddRem} & {\sf Injective} & {\sf Empty[SO]} & {\sf Empty[HB]} & {\sf FIFO-1/LIFO-1/Exchange} & {\sf FIFO-2/LIFO-2} & {\bf Max Time (s)} \\ \hline
{\sf 2Lock} & {\sf MW+MR} & {\sf MW+MR} & {\sf CC} & {\sf CC} & {\sf MW+MR} & {\sf MW+MR} & 269\\ 
{\sf Queue \cite{MS96}} & & {\sf +WFR} & & & & &\\ \hline
{\sf LockFree} & {\sf MW+MR} & {\sf EC} & {\sf CC} & {\sf CC} & {\sf MW+MR} & {\sf EC} & 152\\
{\sf Queue \cite{MS96}} & & &  &  & & & \\ \hline
{\sf HW} & {\sf EC} & {\sf EC} & {\sf RMW} & {\sf MW+MR} & {\sf CC} & {\sf MW+MR} & 61\\ 
{\sf Queue \cite{HW90}} & & & & {\sf +RMW} & & & \\ \hline
{\sf Treiber} & {\sf MW+MR} & {\sf EC} & {\sf CC} & {\sf CC} & {\sf MW+MR} & {\sf EC} & 245\\
{\sf Stack \cite{TR86}} & {\sf +WFR} & & & & {\sf +WFR} & & \\ \hline
{\sf Elimination} & {\sf MW+MR} & {\sf EC} & {\sf CC} & {\sf CC} & {\sf MW+MR} & {\sf MW} & 65\\
{\sf Stack \cite{HSY04}} & {\sf +WFR} & & & &{\sf +WFR} & & \\ \hline
{\sf Exchanger \cite{HSY04}} & {\sf MW} & {\sf EC} & -NA- & -NA- & {\sf MW} & -NA- & 40\\ \hline 
\end{tabular}
\caption{Consistency policies required for various implementations and specifications.}
\label{tab:exp}
\vspace*{-.4in}
\end{table}

We have implemented our bounded verification procedure and applied it
to a number of library implementations that have been widely-used 
in the world of shared-memory systems. We generate FOL formulae for
each implementation as described in \S 4 and dispatch them to Z3 to
determine their satisfiability. For queues, we have used the
{\sf 2LockQueue}, {\sf LockFree Queue} and {\sf Herlihy and Wing (HW)} Queue implementations,
while for stacks, we have applied our approach on the {\sf Treiber}
and {\sf Elimination} Stack implementations. The {\sf Elimination} stack uses
the exchanger implementation, and so we have also checked the
correctness of the exchanger.

Since our analysis takes as input the bound on the number of
invocations ($k$), the consistency policy, and the specification, we
deploy the system as follows: For each implementation and
specification pairing, we start with bound $k = 2$ and the weakest
consistency policy ({\sf EC}). If we do not find any violation, then
we increase the bound by 1 and perform the analysis again.  On the
other hand, if we do find a violation, then by Theorem-1, we know that
it is guaranteed to be an actual violation. We record its structure
from the satisfiable model returned by Z3, and then increase the
consistency policy to the next higher level. We continue this process
until we exhaust our verification time budget (of 1 hour per benchmark
implementation). Note that all the consistency policies that we
consider can be arranged in a lattice \cite{SI15} whereby the higher
one goes up the lattice, the consistency policies become stronger,
which means they allow only a subset of executions that are allowed by
policies weaker than them. Our tool automatically traverses this
lattice to find the weakest consistency policy at which no bounded
violation is found.


Table \ref{tab:exp} summarizes the results of this process. For each
pair of benchmark implementation and correctness specification, it
shows the weakest consistency policy at which we did not find any
violations. This means that at every consistency policy weaker than
the one specified in the table, violations were discovered.  For each
benchmark, we also note the maximum time needed to find a violation
for any specification by Z3.  Some specifications were discussed in \S
3.4, with {\sf Empty[SO]} meaning we replace the relation
$\F{hb}_{\Gamma}$ with $\F{so}_{\Gamma}$ in the specification; the
correctness specifications for {\sf Queues} are given in the Appendix.
Across all benchmarks, we found that the longest history which
violated any specification within the time bound considered consisted
of 6 invocations.

To empirically validate our results, we also executed all the
benchmarks at the appropriate consistency levels on Cassandra, a
real-world replicated data store. We configured Cassandra with 3
replicas running on Amazon EC2 instances at different physical
locations (all on the US East Coast). We randomly generated client
invocations at all 3 replicas and ran each implementation for 4 hours
(on average 92000 invocations/benchmark).  We collected the resulting
traces and checked the specifications. We did not find any violation
of the specifications, and surmise that violations, when they do
occur, manifest in smaller executions that can be systematically
checked by our analysis.

The results yield a number of interesting observations. First and
foremost, note that even for the same benchmark, different correctness
specifications require different consistency policies, ranging from
the weakest, \emph{Eventual Consistency}, ({\sf EC}) to the strongest,
\emph{Causal Consistency}, ({\sf CC}). This suggests that depending
upon the requirements of the clients of the library, there is a
trade-off between consistency and correctness that can be effectively
explored. It has long been known that \emph{Causal Consistency} incurs
a performance penalty \cite{BFGHS12} due to expensive dependency
tracking, significant metadata storage, and long wait times for all
causally dependent data to arrive. A number of recent approaches
\cite{MLC17,BRR17,DGWZ18} have looked at improving the performance of
\emph{Causal Consistency}, mainly by reducing the amount of dependent
data required. Our experiments suggest that many important correctness
properties of library implementations may not require {\sf CC}, but
would work correctly under weaker session guarantees or even {\sf
  EC}. Note that as we discussed in \S 2, {\sf MW+MR} only require all
data to be propagated from the same session, while {\sf MW+MR+WFR}
requires data to be propagated across the entire causal chain.

Another interesting observation is that important properties such as
{\sf Injective} and {\sf FIFO/LIFO} only require {\sf EC} for most
benchmarks. We also notice that for the same correctness
specification, different benchmarks require different consistency
policies, especially among the various {\sf Queue} benchmarks. This
illustrates that clients have flexibility in choosing an
implementation, based on the properties that they need. For example,
an {\sf HW queue} can satisfy the {\sf AddRem} specification at the
weakest consistency policy ({\sf EC}), but requires {\sf CC} for {\sf
  FIFO-1}, which can be satisfied using just session guarantees by
both {\sf 2LockQueue} and {\sf LockFreeQueue}. No single queue
implementation provides all correctness guarantees at the weakest
consistency level.  For stacks, the {\sf Elimination Stack} and the {\sf Treiber Stack} require the same consistency policies for every specification except {\sf LIFO-2}, for which the {\sf Eliminiation Stack} requires {\sf MW} for the {\sf Exchange} property of the underlying {\sf Exchanger} to be satisfied. By analyzing violations, we also found that both the
access pattern of different implementations as well as the semantics
of the data structure (stack vs. queue) played a major role in
determining how and if violations occur.

Note that even though we unroll loops upto a fixed bound, for all
benchmarks except {\sf LockFree Queue}, the unrolling factor does not
matter because in every loop, every iteration except the last only performs
read events, and the values read are only used in the same
iteration. Hence, only the last iteration which performs a
write/update event is relevant; unrolling the loop once is
sufficient.

\begin{figure}
\vspace*{-.3in}
\small
$\xymatrix@R=4pt{
\underline{\textbf{push(1)}} & \underline{\textbf{push}(3)} & \underline{\textbf{pop}:0}\\
\C{5}:\F{U}(\C{Top}, \C{NULL}, \F{L}_1) & \C{5}:\F{U}(\C{Top}, \F{L}_2, \F{L}_3) & \C{6}:\F{R}(\C{Top}, \F{L}_2)\\
\underline{\textbf{push}(2)} & \underline{\textbf{pop}:3} & \hspace*{6pt}\C{7}:\F{R}(\F{L}_2.\C{Val}, 0)\\
\hspace*{-8pt}\C{5}:\F{U}(\C{Top}, \F{L}_1, \F{L}_2) & \C{9}:\F{U}(\C{Top}, \F{L}_3, \F{L}_2) & \hspace*{12pt}\C{9}:\F{U}(\C{Top}, \F{L_2}, \F{L}_1)\\
\underline{\textbf{pop}:1}\\
\hspace*{-20pt}\C{6}:\F{R}(\C{Top}, \F{L}_1)
}$
\caption{A Violation of $\F{LIFO-1}$ by Treiber Stack under MW+MR involving 6 invocations}
\label{fig:ex3}
\vspace*{-.1in}
\end{figure}

In order to illustrate the complex violations automatically generated
by our framework, consider the violation of {\sf LIFO-1} in the
{\sf Treiber} stack implementation under {\sf MW+MR} in Fig
\ref{fig:ex3}. Here, invocations in the same column are in the same
session. Following the notation as used in the specification in \S
3.4, $\gamma_1 = \C{push}(2)$, $\gamma_2 = \C{push}(1)$, $\gamma_3 =
\C{pop}:1$. As a concrete violation of the specification, $\gamma_2$
happens before $\gamma_1$, but $\gamma_3$ returns the value pushed by
$\gamma_2$ even though $\gamma_1$ is unmatched, thus disobeying the
LIFO property.  The reason behind this violation is that another pop
operation (\C{pop}:0) is actually popping the element pushed by
\C{push}(2), but it does not read the value 1 and instead reads the
initial value 0 (thus also violating {\sf AddRem}). As a result, the
last pop operation in the leftmost session sees only the element 1 on
the stack. We note that there is no violation of smaller length under
{\sf MW+MR}. By upgrading the consistency level to {\sf MW+MW+WFR},
the violation is eliminated.

\section{Related Work and Conclusion}
\label{sec:related}

\vspace*{-8pt}

Verifying applications under weak consistency has received significant
attention in recent years. A number of efforts
\cite{BA14,SI15,GO16,KA18,HO19} have looked at the problem of
verifying arbitrary safety invariants while others have considered
verification with respect to distributed database applications and
specific high-level transactional
properties~\cite{BG16,BR18,NJ18,BBE19a,BBE19b,RNDJ19}.  These results are
orthogonal to the work described here, since neither consider the
question of safely migrating performant concurrent libraries to a
replicated environment.

More directly related are proposals to deal with the specification and
verification of various properties of CRDTs
\cite{BU14,ZE14,GO17,NJ19,WEMP19}. CRDTs also offer a library
interface to clients and have been implemented for various data
structures such as set, list, map, etc. They follow a different system
model than the library implementations that we have considered in our
work, and typically do not require any form of synchronization.
However, this requirement imposes stringent constraints on their
design (for example, in an op-based CRDT, all operations have to
commute with each other). We are not aware of any CRDT-like
implementation of concurrent data structures such as {\sf Queue}, {\sf
  Stack} and {\sf Exchangers} that we have considered here.


Prior works \cite{GO17,NJ19} have also developed automated or
semi-automated approaches to verify the convergence of CRDTs, an
important but fairly low-level property that does not shed much
insight on the correctness of libraries built using them.  High-level
correctness specifications of CRDTs are either given in terms of
abstract RDT specifications \cite{BU14,ZE14} or customized
specification frameworks such as replication-aware linearizability
\cite{WEMP19}. Both of these specification styles base their
correctness criteria in terms of a sequential execution of some
reference implementation. However, direct linearization of all
operations in a concurrent execution is not possible in a distributed
environment, and hence both approaches allow relaxations to help
decide a linearization order.  These relaxations typically take the
form of allowing different per-invocation linearizations based on the
type of the invocation and the visibility relation.  This can lead to
complicated specifications that can be substantially different from
their shared-memory counterparts, complicating verification.
 In contrast, we take axiomatic specifications of data
 structures which bear close resemblance to linearizability, and
 seamlessly adapt them to the world of replicated systems with minimal
 changes.  Further, our axiomatic style also allows clients of the
 library to know exactly how the relaxations in a replicated enviroment
 will impact observable behavior. Finally, unlike other prior work, we
 develop a fully automated approach for bounded verification of library
 implementations.

There has also been recent interest in specifying and verifying
library implementations in the context of weak memory models
\cite{DDWD18,RDR19}. While the specification style of weak memory
models bears some superficial resemblance to that of weak consistency,
the underlying system model is quite different, and weak consistency
models in general allow more relaxed behaviors as well as more
fine-grained control than possible under weak memory given their
ability to provide session-level as well as system-wide consistency
guarantees~\cite{BU14}. \cite{RDR19} also proposes axiomatic specifications
of libraries using happens-before and program orders. Our specifications, while
similar in spirit, are more fine-grained and better suited to replicated systems.

To conclude, we tackle the problem of migrating concurrent library
implementations from shared-memory systems to replicated, distributed
ones.  We define a sensible semantics for such implementations on a
replicated store parametric in the consistency policy of the store and
describe how to migrate the correctness specifications for such
libraries with minimal changes. Our bounded verification framework automatically
finds bounded violations of these specifications.  Parametericity of
consistency policies in the analysis allows allows us to find the
weakest policy that eliminates a discovered violation. Our
experiments have demonstrated that the proposed framework is effective in finding non-trivial
violations in a number of challenging and diverse benchmarks. We also find that
that the spectrum of weak consistency policies in replicated
systems can be effectively explored to tradeoff correctness and
performance.

\bibliographystyle{splncs04}
\bibliography{db}
\appendix
\section{Semantics}

\subsection{Language Semantics}
Below, we present all rules of the operational semantics of the language (continuing from \S 3.2).

\rulelabel{L-Invoke}\\
$$
\RULE{\C{c}(s) \equiv \epsilon \quad h'(s) = m(n) \cdot h'' }{(h',\mu,\C{c}, \alpha) \rightarrow (h', \mu[s \rightarrow \mu(s)[\C{a} \rightarrow n]], \C{c}[s \rightarrow I(m)], \alpha)}
$$

Note that the special variable $\C{a}$ designates the argument, and hence is updated in the local variable store of session $s$ in the above rule.

\rulelabel{L-Read}\\
$$
\RULE{\C{c}(s) \equiv \C{v}=\C{l};\C{c}' \quad \sigma = (i,s,\F{R}(l,n)) }{(h',\mu,c,\alpha) \xrightarrow{\sigma} (h', \mu[s \rightarrow \mu(s)[\C{v} \rightarrow n]], \C{c}[s \rightarrow \C{c}'],\alpha)}
$$

\rulelabel{L-Write}\\
$$
\RULE{\C{c}(s) \equiv \C{l}=e;c' \quad \llbracket e \rrbracket_{\mu(s)} = n \quad \sigma = (i,s,\F{W}(\C{l},n)) }{(h',\mu,\C{c},\alpha) \xrightarrow{\sigma} (h', \mu, \C{c}[s \rightarrow c'],\alpha)}
$$

\rulelabel{L-CAS-Fail}\\
$$
\RULE{\C{c}(s) \equiv v=CAS(l,e_1,e_2);c' \quad \sigma = (i,s,\F{R}(l,m)) \quad \llbracket e_1 \rrbracket_{\mu(s)} \neq m }{(h',\mu,\C{c},\alpha) \xrightarrow{\sigma} (h', \mu[s \rightarrow \mu(s)[v \rightarrow \F{False}]], \C{c}[s \rightarrow c'],\alpha)}
$$

\rulelabel{L-CAS-Success}\\
$$
\RULE{\C{c}(s) \equiv v=CAS(l,e_1,e_2);c' \quad \llbracket e_1 \rrbracket_{\mu(s)} = m \quad \llbracket e_2 \rrbracket_{\mu(s)} = n \quad \sigma = (i,s,\F{U}(l,m,n)) }{(h',\mu,\C{c},\alpha) \xrightarrow{\sigma} (h',  \mu[s \rightarrow \mu(s)[v \rightarrow \F{True}]] , \C{c}[s \rightarrow c'])}
$$

\rulelabel{L-Return-Val}\\
$$\RULE{\C{c}(s) \equiv \C{return } e;c' \quad h'(s) = m(k) \cdot h'' \quad \llbracket e \rrbracket_{\mu(s)} = n \\
\alpha = (\Gamma, \F{so}_{\Gamma}) \quad \gamma = (i,m,k,n,s) \quad \alpha' = (\Gamma \cup \{\gamma\}, \F{so}_{\Gamma} \cup \Gamma^{s} \times \{\gamma\} )}{(h', \mu, \C{c},\alpha) \rightarrow (h'[s \rightarrow h''], \mu, \C{c}[s \rightarrow \epsilon],\alpha')}$$

\rulelabel{L-Return}\\
$$
\RULE{\C{c}(s) \equiv \C{return };c' \quad h'(s) = m(k) \cdot h'' \\
\alpha = (\Gamma, \F{so}_{\Gamma}) \quad \gamma = (i,m,k,\bot,s) \quad \alpha' = (\Gamma \cup \{\gamma\}, \F{so}_{\Gamma} \cup \Gamma^{s} \times \{\gamma\} )}{(h', \mu, \C{c}, \alpha) \rightarrow (h'[s \rightarrow h''], \mu, \C{c}[s \rightarrow \epsilon], \alpha')}
$$

\rulelabel{L-If-True}\\
$$
\RULE{\C{c}(s) \equiv \C{if}(b) then c^t else c^f;c' \quad \llbracket b \rrbracket_{\mu(s)} = \F{True}}{(h', \mu, \C{c}, \alpha) \rightarrow (h', \mu, \C{c}[s \rightarrow c^t;c'],\alpha)}
$$

\rulelabel{L-If-False}\\
$$
\RULE{\C{c}(s) \equiv \C{if}(b) then c^t else c^f;c' \quad \llbracket b \rrbracket_{\mu(s)} = \F{False}}{(h', \mu, \C{c}, \alpha) \rightarrow (h', \mu, \C{c}[s \rightarrow c^f;c'],\alpha)}
$$
\rulelabel{L-While-True}\\
$$
\RULE{\C{c}(s) \equiv \C{while}(b)\ \C{do} c^b\ \C{end};c' \llbracket b \rrbracket_{\mu(s)} = \F{True} }{(h', \mu, \C{c}, \alpha) \rightarrow (h', \mu, \C{c}[s \rightarrow c^b;\C{c}(s)], \alpha)}
$$
\rulelabel{L-While-False}\\
$$
\RULE{\C{c}(s) \equiv \C{while}(b)\ \C{do} c^b\ \C{end};c' \llbracket b \rrbracket_{\mu(s)} = \F{False} }{(h', \mu, \C{c}, \alpha) \rightarrow (h', \mu, \C{c}[s \rightarrow c'], \alpha)}
$$

\subsection{Replicated Store Semantics}

\rulelabel{R-Read}\\
$$
\RULE{\Sigma' \subseteq \Sigma^l_W \cup \Sigma^l_U \quad \sigma' \in \F{MAX}_{\F{ar}}(\Sigma') \\
act(\sigma') = \F{W}(l,n) \vee act(\sigma') = \F{U}(l,\_,n) \quad \sigma = (i,s,R(l,n))\\
\F{vis}' = \F{vis} \cup \Sigma' \times \{\sigma\} \quad \F{so}' = \F{so} \cup \Sigma^s \times \{\sigma\} \quad \psi(\Sigma, \F{vis}', \F{ar}, \F{so}')}{
(\Sigma, \F{vis}, \F{ar}, \F{so}) \xrightarrow{\sigma} (\Sigma \cup \{\sigma\},\F{vis}',\F{ar},\F{so}' )}
$$

\rulelabel{R-Write}\\
$$
\RULE{\sigma = (i,s,W(l,n)) \quad \F{ar} \subseteq \F{ar}' \quad  \F{ar}'\ \F{is}\ \F{a}\ \F{total}\ \F{order}\ \F{on}\ \Sigma^l \cup \{\sigma\} \\ \forall \sigma_1,\sigma_2. \neg(\F{hb}(\sigma_1, \sigma_2) \wedge \F{ar}'(\sigma_2, \sigma_1))  \quad so' = so \cup \Sigma^s \times \{\sigma\}}{
(\Sigma, \F{vis}, \F{ar}, \F{so}) \xrightarrow{\sigma} (\Sigma \cup \{\sigma\},\F{vis},\F{ar}',\F{so}' )}
$$

\subsection{Correctness Specification}
We now provide the various correctness axioms that we use for the Queue and Exchanger data structures. The Queue data structure has two methods $M = \{\C{Enqueue},\C{Dequeue}\}$. We define the $\F{match}$ predicate as follows:
\[\F{match}(\gamma_1, \gamma_2) \Leftrightarrow \F{m}(\gamma_1) = \C{Enqueue} \wedge \F{m}(\gamma_2) = \C{Dequeue} \wedge \F{a}(\gamma_1) = \F{r}(\gamma_2)\]

The axioms for Queue are as follows:
\begin{itemize}
\item $\F{AddRem}$ : $\forall \gamma \in \Gamma. \F{m}(\gamma) = \C{Dequeue} \wedge \F{r}(\gamma) \neq \C{EMPTY} \Rightarrow \exists \gamma' \in \Gamma. \F{match}(\gamma', \gamma)$
\item $\F{Injective}$ : $\forall \gamma_1, \gamma_2, \gamma_3 \in \Gamma. \F{match}(\gamma_1, \gamma_2) \wedge \F{match}(\gamma_1, \gamma_3) \Rightarrow \gamma_2 = \gamma_3$
\item $\F{Empty}$ : $\forall \gamma_1, \gamma_2, \gamma_3 \in \Gamma. \F{m}(\gamma_1) = \C{Dequeue} \wedge \F{r}(\gamma_1) = \C{EMPTY} \wedge \F{m}(\gamma_2) = \C{Enqueue} \wedge \F{hb}_{\Gamma}(\gamma_2, \gamma_1) \Rightarrow \exists \gamma_3 \in \Gamma. \F{match}(\gamma_2, \gamma_3)$
\item $\F{FIFO-1}$ : $\forall \gamma_1, \gamma_2, \gamma_3 \in \Gamma. \F{m}(\gamma_1) = \C{Enqueue} \wedge \F{match}(\gamma_2, \gamma_3) \wedge \F{hb}(\gamma_1, \gamma_2) \Rightarrow \exists \gamma_4 \in \Gamma. \F{match}(\gamma_1, \gamma_4)$
\item $\F{LIFO-2}$ : $\forall \gamma_1, \gamma_2, \gamma_3, \gamma_4 \in \Gamma. \neg (\F{match}(\gamma_1, \gamma_4) \wedge \F{match}(\gamma_2, \gamma_3) \wedge \F{hb}(\gamma_1, \gamma_2) \wedge \F{hb}(\gamma_3, \gamma_4))$
\end{itemize}

The Exchanger data structure has one method $M = \{\C{Exchange}\}$. The \C{Exchange} method takes as argument the value to be exchanged, and returns the value from its pairing \C{Exchange} method. If the exchange fails, then it returns the value $\bot$. The axioms for Exchanger are defined as follows:
\begin{itemize}
\item $\F{AddRem}$ : $\forall \gamma_1 \in \Gamma. \F{r}(\gamma_1) \neq \bot \Rightarrow \exists \gamma_2. \F{a}(\gamma_2) = \F{r}(\gamma_1)$
\item $\F{Injective}$ : $\forall \gamma_1,\gamma_2,\gamma_3 \in \Gamma. (\F{r}(\gamma_2) = \F{a}(\gamma_1) \wedge \F{r}(\gamma_3) = \F{a}(\gamma_1)) \Rightarrow \gamma_2 = \gamma_3$
\item $\F{Exchange}$ : $\forall \gamma_1,\gamma_2 \in \Gamma. \F{a}(\gamma_1) = \F{r}(\gamma_2) \Rightarrow \F{r}(\gamma_1) = \F{a}(\gamma_2)$
\end{itemize}

\section{Bounded Verification}

Here, we specify the rules for generating constraints for write and CAS statements in the implementation:

$$
\RULE{\F{Impl}(m):n:\ \C{l} = \C{e}}{\forall i \in \invo.\ (\F{meth}(i) = m \wedge \llbracket \phi_{mn} \rrbracket_{i}) \Rightarrow (\F{Etype}(\F{P}_{mn}(i)) = \F{W} \wedge \F{loc}(\F{P}_{mn}(i)) = \C{l} \\ \wedge \F{wval}(\F{P}_{mn}(i)) = \llbracket e \rrbracket_i)}
$$

Note that $\llbracket e \rrbracket_i$ replaces every occurence of amy local variable \C{v} with its projection function $\rho_{\C{v}}$.

$$
\RULE{\F{Impl}(m):n:\ \C{v} = \C{CAS}(\C{l}, e_1, e_2)}{\forall i \in \invo.\ (\F{meth}(i) = m \wedge \llbracket \phi_{mn} \rrbracket_{i}) \Rightarrow (\F{P}_{mn}(i) \neq \bot \wedge \\ \F{loc}(\F{P}_{mn}(i)) = \C{l})\\
\forall i \in \invo.\ (\F{meth}(i) = m \wedge \llbracket \phi_{mn} \rrbracket_{i} \wedge \F{rval}(\F{P}_{mn}(i)) \neq \llbracket e_1 \rrbracket_{i}) \\ \Rightarrow \F{Etype}(\F{P}_{mn}(i)) = \F{R}\\
\forall i \in \invo.\ (\F{meth}(i) = m \wedge \llbracket \phi_{mn} \rrbracket_{i} \wedge \F{rval}(\F{P}_{mn}(i)) = \llbracket e_1 \rrbracket_{i}) \Rightarrow (\F{Etype}(\F{P}_{mn}(i)) = \F{U} \wedge \\ \F{wval}(\F{P}_{mn}(i) = \llbracket e_2 \rrbracket_i)
}
$$

\section{Proof of Theroem 1}
\textbf{Theorem 1.} Given a library implementation $L$, consistency policy $\Psi$ and a
correctness axiom $\theta$, if the collection of formulae described
above are satisfiable, then there exists a history $h$ and an abstract
execution $\Gamma \in \llbracket \Omega_{h,L,\Psi} \rrbracket$ which
violates $\theta$.
\begin{proof}
We provide a proof sketch. Given the satisfiable model of the collection of formulae, from the functions $\F{meth}$, $\F{arg}$, $\F{sess}$, and the predicate $\F{so}_{\invo}$, we can construct the history $h$ which would consist of the invocations corresponding to each $\F{INV}_i$. From the different elements of the session sort $\F{S}$, we obtain the set of sessions $S$. We can also construct the abstract execution $\alpha$ of $h$ that violates the specification, by additionally using the function $\F{ret}$. Since the specification $\theta$ only concerns the method type, argument value, return value and session order relation among invocations, and since $\neg \theta$ is part of the formulae, it is clear that $\alpha$ does actually violate the specification. The only thing left to show now is that $\alpha$ is also a valid abstract execution, i.e. $\alpha \in \llbracket \Omega_{h,L,\Psi} \rrbracket$.

To show this, we use all the events $\F{P}_{mn}(i)$ from the satisfiable model for all invocations $i$. In particular, when $\F{P}_{mn}(i) \neq \bot$, then the generated event will be a label produced during the sequence of transitions of $\Omega_{h,L,\Psi}$ leading to $\alpha$. In order to create this sequence of transitions, we will use the predicates $\F{vis}$ and $\F{so}_{\event}$. In particular, we consider the transitive closure of the union of these predicates, $(\F{vis} \cup \F{so}_{\event})^{*}$, and then linearize it (i.e. add more orderings) to make it a total order $\F{hb}_{\event}$. Given an event $e = \F{P}_{mn}(i)$, we can determine the invocation ($i$) and the particular statement ($n$) in the implementation generating this event. We now traverse through all events $\F{P}_{mn}(i)$ in the order $\F{hb}_{\event}$ and use the corresponding language semantics rules and replicated store semantics rules to generate the transition sequence. Since all the pre-conditions of both the language semantics and the replicated store semantics rules are encoded in the formulae and hence are satisfied, the events will have the correct read values and write values concurring with the functions $\F{rval}$ and $\F{wval}$. Further, the pre-conditions of abstract execution rules are also satisfied, and hence, the return values of the invocations added to the abstract execution will also concur with the function $\F{ret}$. This implies that the abstract execution $\alpha$ will be generated at the end of the sequence of transitions after exhausting all the events $\F{P}_{mn}(i)$, and as we saw earlier, $\alpha$ does violate the specification $\theta$.
\end{proof}

\end{document}